\documentclass[10pt]{article}
\usepackage[body={17cm, 23cm, centered}]{geometry}

\usepackage{amsmath,amssymb,amsfonts}
\usepackage{mathtools}
\usepackage{bm}
\usepackage{mathrsfs}
\usepackage{lmodern} 
\usepackage{simpler-wick}

\usepackage{subcaption}
\usepackage[labelfont=bf, margin=20pt, font={small}]{caption}
\usepackage[utf8]{inputenc}
\usepackage[T1]{fontenc}
\usepackage[english]{babel}
\usepackage{graphicx} 
\usepackage{cite}
\usepackage{braket}
\usepackage{textcomp}
\usepackage{slashed}
\usepackage[dvipsnames,svgnames,table]{xcolor}
\usepackage{tikz}
\usepackage{tikz-cd}
\usepackage[normalem]{ulem}
\usepackage{float}
\usepackage{multirow}
\usepackage[makeroom]{cancel}

\graphicspath{{figs/}}

\usepackage[debug,pageanchor=false]{hyperref}
\hypersetup{
    colorlinks=true,
    linktocpage,
    breaklinks,
    urlcolor=blue,
    linkcolor=blue,
    citecolor=blue
}

\newcommand\beq{\begin{equation}}
\newcommand\eeq{\end{equation}}
\newcommand{\pd}{\partial}

\numberwithin{equation}{section}
\begin{document}

\begin{titlepage}
\phantom{preprint}
\vspace{0.5cm}

\begin{center}
   \baselineskip=16pt
   {\large \bf Global quenches and correlator dynamics in de Sitter space
   }
   \vskip 1.5cm
   Ivan A. Belkovich$^{a}$\footnote{\tt belkovich.ia@phystech.edu }, Damir Sadekov$^{a,b}$\footnote{\tt sadekov.di@phystech.edu }, Gleb S. Zverev$^{b}$\footnote{\tt zverev.gs@phystech.edu }
       \vskip .6cm
             \begin{small}
                          {\it
                          $^a$P. N. Lebedev Physical Institute, Moscow 119991, Russia\\
                          $^b$Moscow Institute of Physics and Technology, 
                          Laboratory of High Energy Physics\\
                          Institutskii per., 9, 141702, Dolgoprudny, Russia
                          } \\ 
\end{small}
\end{center}
\vspace{0.5cm}
\begin{center} 
\textbf{Abstract}
\end{center}
We study non-equilibrium initial states of quantum fields in curved space-time and develop a framework for describing global quenches as unitary perturbations of the initial density matrix. Using the Keldysh–Schwinger functional integral, we derive expressions for post-quench correlators in arbitrary geometries and apply the method to both Minkowski and de Sitter backgrounds. In flat space, the approach reproduces the known relaxation behaviour for massive fields and reveals qualitatively distinct dynamics in the massless case. In de Sitter space, we find that the late-time decay of quench-induced corrections depends sensitively on the mass and may influence the behaviour of cosmological observables such as the power spectrum. The framework also extends straightforwardly to non-Gaussian initial states, providing a basis for future studies of loop effects and primordial non-Gaussianities.

\vfill
\setcounter{footnote}{0}
\end{titlepage}

\newpage

\tableofcontents

\section{Introduction}
Non-equilibrium processes in quantum systems attract significant attention both in condensed matter physics and in the context of quantum field theory and holography. One of the most universal and theoretically controlled ways to drive a system out of equilibrium is a quantum quench, usually defined as a sudden change of some Hamiltonian parameter, such as the mass or coupling constant. Studying the subsequent relaxation allows one to explore fundamental questions of thermalization, integrability, and the evolution of quantum correlations. In this context one usually distinguishes between the so-called local quench, where the system is perturbed in a spatially localized region, and the global quench, where the sudden change is uniform throughout space. Local quenches are actively studied in condensed matter physics \cite{stephan2011local, PhysRevLett.108.077206, PhysRevLett.121.030601}, as well as in the investigation of chaotic and integrable features in different systems, including holographic models \cite{haldar2021signatures, he2013initial, Modak:2020faf, Calabrese:2007mtj, Ageev:2025iiy, Ageev:2025yiq, Ageev:2018nye, Ageev:2023gna}. Global quenches are typically considered in the context of relaxation toward some equilibrium state, commonly referred to as ``subsystem thermalization'' \cite{Calabrese:2006rx, Sotiriadis:2010si, Banerjee:2019ilw, Calabrese:2007rg, Das:2014hqa, Calabrese:2016xau, Ageev:2023wrb}. They are also a standard tool for analyzing the dynamics of phase transitions \cite{Hung:2012zr, chandran2012kibble, Das:2014hqa} and the evolution of quantum entanglement following a sudden perturbation \cite{Calabrese:2016xau, Moosa:2017yvt}.

In the simplest setting, a global quench is realized by a mass-dependent $m(t)$ quench protocol in the sudden limit \cite{Sotiriadis:2010si, Banerjee:2019ilw}, where the mass instantaneously changes from one value to another, $m\rightarrow M$. For flat spacetime, in free theories as well as certain interacting and conformal theories, the dynamics of such quenches is well understood when the initial state is either the vacuum of the pre-quench theory or specific squeezed states. In this situation, the post-quench state can be written as a generalized Cardy-Calabrese (gCC) state, and the system relaxes toward a generalized Gibbs ensemble, with the relaxation rate depending on the number of spatial dimensions \cite{Sotiriadis:2010si, Banerjee:2019ilw}.

In recent decades, it has been recognized that mechanisms analogous to global quenches may also play a role in the evolution of the early Universe. In this context, a global quantum quench can effectively describe processes experienced by fields after fast transient phenomena, such as phase transitions induced by a sharp change in temperature, or sudden features in the inflaton potential \cite{Boyanovsky:1996rw, Boyanovsky:1997xt, Boyanovsky:2006bf, Carrilho:2016con, Joy:2007na, Hazra:2010ve,  Banerjee:2021lqu}. Most existing works focus on situations involving a change of phase \cite{Boyanovsky:1996rw, Boyanovsky:1997xt, Boyanovsky:2006bf}, but it is also of interest to analyze non-equilibrium dynamics triggered by perturbations that do not induce a phase transition \cite{Carrilho:2016con, Banerjee:2021lqu}. More generally, a quench may serve as a model for restructuring the initial quantum state during or shortly after inflation. For example, the mass-dependent quench protocol in the sudden limit for the inflaton potential was used in \cite{Joy:2007na} to explain observable features in the scalar primordial power spectrum. In \cite{Carrilho:2016con}, a quench was analyzed in the $O(N)$ model in de Sitter spacetime, where it was shown that after a sudden mass shift, the effective mass dynamically returns to some asymptotic value, although at intermediate stages its square may become negative, corresponding to a temporary local spontaneous symmetry breaking. An interesting approach was suggested in \cite{Banerjee:2021lqu}, where a generalization of the Caldeira-Leggett model to curved spacetime produced a mass-dependent quench protocol as an effective description after integrating out environmental degrees of freedom. In this case, the sudden mass quench arises naturally, and both the vacuum state and Gaussian squeezed states were considered as initial states. It was shown that after relaxation the system approaches a gCC-like state in de Sitter space, and the analysis of two-point correlators allowed one to describe the corresponding modifications of the perturbation spectrum. A closely related setup was studied in \cite{Albrecht:2014hxa}, where flat spacetime is suddenly transformed into de Sitter spacetime. It was demonstrated that the evolution of local operators, such as the stress-energy tensor, approaches the Bunch-Davies state at late times even if the initial state breaks de Sitter symmetry, allowing one to regard it as a universal attractor. This raises the question of how much memory of the initial state is retained in quantum dynamics in an expanding geometry, and what the nature of relaxation is in curved spacetime.

We note that gCC-type states are always Gaussian and non-unitary constructions, which nevertheless allow for analytic treatment of correlators in globally out-of-equilibrium systems. In this work, we propose another class of initial states suitable for describing post-quench dynamics in quantum field theory in curved spacetime. We use the Keldysh functional integral formalism \cite{Berges:2004yj, kamenev2023field, Arseev:2015, Akhmedov:2013vka} to derive explicit expressions for two-point correlation functions after a global (in space) perturbation implemented as a unitary transformation of the equilibrium density matrix (see Appendix \ref{app:keldysh_formalism}). The idea of this method was introduced in \cite{Radovskaya:2023ctz}, and in the case of a local quench in flat spacetime it reproduces known results \cite{Radovskaya:2023ctz, Ageev:2022kpm}. We find that for Gaussian perturbations the resulting behavior qualitatively matches that of the sudden mass quench protocol for massive fields (see Appendix \ref{app:flat}). At the same time, for massless fields in flat spacetime we observe the absence of relaxation even in $D=4$, which is the feature of the behavior known for $D=2,3$ in the $m(t)$ quench protocol.

More precisely, we consider a theory of a scalar field $\phi$ and perturb the density matrix of the pure vacuum state $\widehat{\rho}_0$ as $\widehat{\rho} = \widehat{Q}\widehat{\rho}_0\widehat{Q}^{\dagger}$, where $\widehat{Q} = \exp\left\{-i\alpha\int d^{D-1}\bm{x}\sqrt{|g|}\;\widehat{\phi}^2(\bm{x})\right\}$ with $\alpha$ an arbitrary mass-dimensional parameter $[\alpha]=[m]$. Then the evolution of the Keldysh two-point function $G_{K}^{Q}$ after the quench can be expressed in terms of integrals over the equilibrium Keldysh function $G_{K}$ and retarded Green’s function $G_R$ (Appendix \ref{app:keldysh_formalism}, \cite{Radovskaya:2023ctz}):
\beq\label{eq:Keldysh_evolution_general}
\begin{aligned}
    iG_{K}^{Q}(t,\bm{x}|t,\bm{y}) =
    \\
    = iG_K(t,\bm{x}|t,\bm{y}) - 2\alpha\int d^{D-1}\bm{\xi}\sqrt{|g(t_0,\bm{\xi})|}\bigg[iG_K(t,\bm{x}|t_0,\bm{\xi})G_R(t,\bm{y}|t_0,\bm{\xi}) + iG_{K}(t,\bm{y}|t_0,\bm{\xi})G_R(t,\bm{x}|t_0,\bm{\xi}) \bigg] 
    \\
    +\; 4\alpha^2 \int d^{D-1}\bm{\xi}_1 d^{D-1}\bm{\xi}_2 \sqrt{|g(t_0,\bm{\xi}_1)|}\sqrt{|g(t_0,\bm{\xi}_2)|}\;iG_K(t_0,\bm{\xi}_1|t_0,\bm{\xi}_2)G_R(t,\bm{x}|t_0,\bm{\xi}_1)G_R(t,\bm{x}|t_0,\bm{\xi}_2)
\end{aligned}
\eeq
or, in momentum space (assuming $g_{\mu\nu}$ depends only on time and spatial slices are homogeneous)
\beq\label{eq:keldysh_evolution_momentum_space}
\begin{aligned}
    iG_{K}^{Q}(t,\bm{x}|t,\bm{y}) = \int\frac{d^{D-1}\bm{k}}{(2\pi)^{D-1}}\;e^{i\bm{k}(\bm{x-y})}\;\bigg\{iG_K(\bm{k}|t,t) \;-\;4\alpha\sqrt{|g(t_0)|}\;iG_{K}(\bm{k}|t,t_0)G_{R}(\bm{k}|t,t_0)\;+
    \\
    +\;4\alpha^2\sqrt{|g(t_0)|}\sqrt{|g(t_0)|}iG_{K}(\bm{k}|t_0,t_0)G_{R}(\bm{k}|t,t_0)G_{R}(\bm{k}|t,t_0)\bigg\}=
    \\
    = iG_{K}(t,\bm{x}|t,\bm{y}) - 4\alpha I_1(t,t_0,r) + 4\alpha^2 I_2(t,t_0,r),
\end{aligned}
\eeq
where $I_1$ and $I_2$ denote the integrals giving the corresponding corrections to the two-point correlator, and we assume the equilibrium state to be homogeneous and isotropic, so the correlators depend only on $r = |\bm{x}-\bm{y}|$. In $D=4$ Minkowski spacetime, for massive fields we find relaxation of the two-point correlator at late times $mt\gg 1$:
\beq\label{eq:post_quench_Keldysh_flat_spacetime_Intro}
\begin{aligned}
    iG_{K}^{Q}(t,\bm{x}|t,\bm{y}) \simeq iG_{K}^{\text{post-quench}}(r) + \frac{\alpha\sqrt{\alpha^2+m^2}}{8\pi^{\frac{3}{2}}}\cdot\frac{1}{(mt)^{\frac{3}{2}}}\cos\Big(2mt+\delta\Big),
    \\
    \delta = \frac{\pi}{4}-\arctan\left(\frac{\alpha}{m}\right),
\end{aligned}
\eeq
where $iG_{K}^{\text{post-quench}}(r)$ is the modified two-point correlator after the quench, and the relaxation rate qualitatively matches the result for the $m(t)$ quench protocol \cite{Sotiriadis:2010si, Banerjee:2019ilw}. However, for massless $m=0$ fields in Minkowski spacetime and $t\gg r$, the correlator does not relax even in four dimensions:
\beq\label{eq:non_relaxation_massless_flat}
    iG^{Q}_{K}\sim\alpha^{2}\log\left[\frac{2t}{r}\right].
\eeq
For mass quenches this behavior appears only in lower dimensions, so a more detailed analysis of critical post-quench dynamics within our framework is left for future work.

In de Sitter spacetime the situation is different: even equilibrium Green’s functions of the free theory decay with time as a power of the conformal time $\eta=e^{-t}$ (taking the Hubble parameter $H=1$). Therefore, regarding relaxation, one first needs to compare the decay of local operators in the non-equilibrium setup with the decay of the correlation function for the Bunch-Davies state. Second, given the secular growth of loop corrections even for the Bunch-Davies state \cite{Akhmedov:2013vka, Krotov:2010ma, Serreau:2013psa, Akhmedov:2019cfd, Akhmedov:2024npw}, it would be interesting to analyze how controlled non-equilibrium perturbations affect long-time dynamics in the presence of growing loop contributions. Understanding the two-point correlator is central to all such questions, and the main goal of this work is to demonstrate the application of our method to these correlation functions in curved spacetime. In Section \ref{sec:EPP}, we analyze the behavior of the Keldysh propagator at equal times after the quench in the expanding Poincar\`e patch of de Sitter space. We find that for short spatial separations $r\lesssim \eta$, the qualitative behavior changes depending on whether the mass is above or below the critical value $m_{cr}=\sqrt{2}$ corresponding to the ``conformally coupled'' scalar. We expect that this may have nontrivial implications for local operators such as the stress-energy tensor, where one takes the coincidence limit $r\rightarrow 0$. However, the situation remains unclear and we leave the full analysis, including a proper regularization procedure for non-equilibrium correlators, for future work. In Section \ref{sec:comment}, we discuss the application of our approach to non-Gaussian perturbations of equilibrium states. In Appendices \ref{app:keldysh_formalism} and \ref{app:flat}, we describe our construction of the perturbed initial state and the corresponding evolution of correlators in curved spacetime, and we also present explicit computations in flat spacetime.

The proposed approach also enables systematic exploration of relaxation for non-Gaussian and anisotropic initial states in arbitrary geometries where Green's functions are known. In particular, the derivation presented in Appendix \ref{app:keldysh_formalism} applies with minimal modification if the parameter $\alpha$ is replaced by an arbitrary anisotropic spatial function $\alpha(\bm{x})$. Additionally, for future studies of loop corrections, we plan to generalize the approach to compute modified Green’s functions at different time arguments $t_1\neq t_2$. We hope that this class of states will broaden the study of undergoing physics in the early Universe, and potentially find applications in the study of chaos, holographic models, and more generally non-equilibrium dynamics across different physical systems.

\section{Quantum quench in the expanding Poincar\`e patch}\label{sec:EPP}
In this section we study the behavior of the two-point correlator of a free minimally coupled scalar field after a Gaussian quench, mentioned above and described in details in Appendix \ref{app:keldysh_formalism}, in the Poincar\`e patch (EPP) of de Sitter space:
\beq\label{eq:action}
    S[\phi] = \int d^{D}x\sqrt{|g|}\left[ \frac{1}{2}g^{\mu\nu}\partial_{\mu}\phi\partial_{\nu}\phi - \frac{1}{2}m^2\phi^2\right],
\end{equation}
\beq\label{eq:metric}
    ds^2 = \frac{1}{\eta^2}\left(d\eta^2 - d\bm{x}^2 \right),
\eeq
where $\eta=e^{-t}$ is conformal time, and we set the Hubble parameter to unity, $H=1$. The canonical quantization of the field is as follows:
\beq \label{eq:EPP_field_decomposition}
    \begin{aligned}
        \widehat{\phi}(\eta,\bm{x})=\int\dfrac{d^{D-1}\bm{k}}{(2\pi)^{D-1}}&\bigg[
\widehat{a}_{\bm{k}}f_{\bm{k}}(\eta)e^{i\bm{kx}}+
\widehat{a}^{\dagger}_{\bm{k}}f^{*}_{\bm{k}}(\eta)e^{-i\bm{kx}}
\bigg], \quad 
\left[\widehat{a}_{\bm{k}},\widehat{a}_{\bm{p}}^\dagger 
\right] = (2\pi)^{D-1}\delta(\bm{k}-\bm{p}),
   \end{aligned}
\eeq
where the harmonics $f_{\bm{k}}(\eta)e^{i\bm{kx}}$ solve the free Klein–Gordon equation for the action (\ref{eq:action}). The solutions are given by Hankel functions, and we fix the modes so that the Bunch–Davies state $\Big|\text{BD}\Big\rangle$ corresponds to the Fock vacuum for the annihilation operators in (\ref{eq:EPP_field_decomposition}):
\beq \label{eq:harmonics}
\begin{aligned}
    f_{\bm{k}}(\eta) = \eta^{\frac{D-1}{2}}h_{\nu}\left(k\eta\right) = \eta^{\frac{D-1}{2}}\frac{\sqrt{\pi}}{2}H_{\nu}^{(1)}\left(k\eta\right),\;\; \nu = \sqrt{\frac{(D-1)^2}{4}-m^2},\;\;\widehat{a}_{\bm{k}}\Big|\text{BD}\Big\rangle = 0.
\end{aligned}
\eeq
In what follows we assume the most interesting case of light fields (complementary series) $m<\frac{D-1}{2}$, and we work only in $D=4$ dimensions. It is convenient to keep in mind the following asymptotic expansions of the Hankel functions:
\beq\label{eq:asymptotics}
\begin{aligned}
    &h_{\nu}(k\eta) \simeq \frac{1}{\sqrt{2 k\eta}} e^{ik\eta}e^{-i\frac{\pi}{2}\left(\nu+\frac{1}{2}\right)}, \quad {\rm as} \quad k\eta \gg |\nu| ;
    \\
    &h_{\nu}(k\eta) \simeq iA_{-}\frac{1}{(k\eta)^{\nu}} + iA_{+}(k\eta)^{\nu} + iB(k\eta)^{-\nu+2} +\ldots, \quad {\rm as} \quad k\eta \ll |\nu|;
    \\
    &{\rm where} \quad A_{-} = -\frac{\Gamma(\nu)}{2\sqrt{\pi}}2^{\nu}, \; A_{+} = \frac{\sqrt{\pi}e^{-i\pi\nu}}{2^{\nu+1}\Gamma(\nu+1)\sin(\pi\nu)}, \; B = \frac{2^\nu}{8\sqrt{\pi}}\frac{\Gamma(\nu)}{2-\nu}.
\end{aligned}
\eeq
Then, for the evolution of the two-point correlator after perturbing the state $\Big|\text{BD}\Big\rangle$ by the Gaussian operator (\ref{eq:density_matrix_change_after_quench}), (\ref{eq:qudratic_quench}), we obtain in the EPP:
\beq\label{eq:EPP_general_keldysh_Q}
\begin{aligned}
    iG_{K}^{Q}(t,\bm{x}|t,\bm{y})
    = iG_{K}(t,\bm{x}|t,\bm{y}) - 4\alpha I_1^{\text{EPP}}(t,t_0,r) + 4\alpha^2 I_2^{\text{EPP}}(t,t_0,r),
\end{aligned}
\eeq
where $r=|\bm{x}-\bm{y}|$ and
\begin{equation}\label{eq:I1_general_EPP}
    \begin{aligned}
        I_1^{\text{EPP}} = -\int\frac{d^{D-1}\mathbf{k}}{(2\pi)^{D-1}}e^{i\mathbf{k}\mathbf{r}}\eta^{D-1}\text{Im}\Big\{h_{\nu}^2(k\eta)h_{\nu}^{*2}(k\eta_0)\Big\} =
        \\
        =\;-\frac{1}{(2\pi)^{\frac{D-1}{2}}r^{\frac{D-3}{2}}}\int_{0}^{\infty}d kk^{\frac{D-1}{2}}J_{\frac{D-3}{2}}(kr)\eta^{D-1} \text{Im}\Big\{h_{\nu}^2(k\eta)h_{\nu}^{*2}(k\eta_0)\Big\}=
        \\
        =-\frac{1}{2\pi^2}\frac{\eta}{r}\int_{0}^{\infty}d\xi\;\xi\cdot\sin\Big(\xi\frac{r}{\eta}\Big)\text{Im}\left\{h_{\nu}^2(\xi)\cdot h_{\nu}^{*2}\left(\xi\frac{\eta_0}{\eta}\right)\right\},
    \end{aligned}
\end{equation}
\begin{equation}\label{eq:I2_general_EPP}
    \begin{aligned}
        I_2^{\text{EPP}} = \frac{2}{\pi^2}\frac{\eta}{r}\int_{0}^{\infty}d\xi\;\xi\cdot\sin\Big(\xi\frac{r}{\eta}\Big)\left|h_{\nu}\left(\xi\frac{\eta_0}{\eta}\right)\right|^2\text{Im}^{2}\left\{h_{\nu}(\xi)\cdot h_{\nu}^{*}\left(\xi\frac{\eta_0}{\eta}\right)\right\}.
    \end{aligned}
\end{equation}
We are particularly interested in the behavior of the Keldysh propagator, since it is sensitive to the evolution of the state of the theory. In the last line of Eq. (\ref{eq:I1_general_EPP}) we set $D=4$. We will estimate the behavior of the two-point correlator at late times $\eta\ll1$ in the regime where the quench happened sufficiently early, $\eta_0\gg1$, and we will also assume that the spatial separation is small, $r=|\bm{x}-\bm{y|}\ll\eta_0$. Let us note immediately that these expressions can be used in other regimes as well, in particular for studying loop corrections, where the integrals over internal vertices run over the whole spacetime. However, the main goal here is to demonstrate a controlled way to drive the system out of equilibrium, so we focus on the simplest case for estimates and leave a more advanced analysis for future work. The integrals (\ref{eq:I1_general_EPP})--(\ref{eq:I2_general_EPP}) can then be estimated using the asymptotics (\ref{eq:asymptotics}) by splitting the $\xi$-integration into the ranges $\left[0,\nu\frac{\eta}{\eta_0}\right)$, $\left[\nu\frac{\eta}{\eta_0},\nu\right]$, and $\left(\nu, \infty\right)$. Replacing the harmonics by the corresponding asymptotics in these regions yields the estimates:
\beq\label{eq:EPP_I1_estimation}
    I_1^{\text{EPP}} \simeq a(r)\cdot \left(\frac{\eta}{\eta_0}\right)^{3-2\nu} + \;\text{subleading terms},
\eeq
\beq\label{eq:EPP_I2_estimation}
\begin{aligned}
    I_2^{\text{EPP}} \simeq  \left(\frac{\eta}{\eta_0}\right)^{3-2\nu}\cdot\Bigg[b\;+\; \frac{A_{-}^2}{4\pi^2}\frac{\eta_0}{r}\int_{\nu}^{\nu\frac{\eta_0}{\eta}}\frac{d\ell}{\ell^{1+2\nu}}\;\sin\left(\ell\frac{r}{\eta_0}\right)\cdot\Big(1+\sin\left(2\ell-\pi\nu\right)\Big)\Bigg]\; -
    \\
    -\;\frac{1}{8\pi^2}\frac{\eta^2}{\eta_0^2}\text{Ci}\left(\nu\frac{r}{\eta}  \right)\;+\;\text{subleading terms},
\end{aligned}
\eeq
where $\text{Ci}$ is the cosine integral and 
\beq\label{eq:EPP_coefficients_I1I2_etimation}
\begin{aligned}
    a(r)\simeq \frac{A_{-}^{3}\text{Im}\Big\{A_+\Big\} }{\pi^2}\frac{\nu^{3-2\nu}}{3-2\nu}\; + \;\frac{A_{-}^2}{4\pi^2}\frac{\eta_0}{r}\int_{\nu}^{\infty}\frac{d\ell}{\ell^{2\nu}}\sin\Big(\ell\frac{r}{\eta_0}\Big)\cos\Big(2\ell-\pi\nu\Big),\quad
    b = \frac{2A_{-}^{4}\text{Im}^{2}\Big\{A_+\Big\} }{\pi^2}\frac{\nu^{3-2\nu}}{3-2\nu}\;.
\end{aligned}
\eeq
Here the expression for the coefficient $a$ is approximate because we extended the upper limit of the integral over $\ell$ to infinity instead of $\nu\frac{\eta_0}{\eta}\gg1$, and the integral is always convergent due to oscillations. It is also important that the term with the cosine integral $\text{Ci}$ in (\ref{eq:EPP_I2_estimation}) is suppressed in the limit $\eta\ll r$ as $\left(\frac{\eta}{\eta_0}\right)^{3}$. Nevertheless, we keep it because for $r\rightarrow0$ it contains a subleading divergence, $\text{Ci}\left(\nu\frac{r}{\eta}\right)\sim\log\left(\frac{r}{\eta}\right)$. While the tree-level two-point correlator diverges quadratically, $iG_{K}\sim\eta^2\big/r^2$, any further study of the evolution of expectation values of local operators, such as e.g. the stress–energy tensor $\langle T_{\mu\nu}\rangle^{Q} \sim \widehat{\mathcal{D}}_{\mu\nu}iG_{K}^{Q}(t,\bm{x}|t,\bm{y})\Big|_{\bm{x}\rightarrow \bm{y}}$ (with some differential operator $\widehat{\mathcal{D}}_{\mu\nu}$), will require careful ultraviolet regularization and subtraction of divergences. 

\begin{figure}[t]
    \centering
    \def\svgwidth{\textwidth}
    \includegraphics[width=\linewidth]{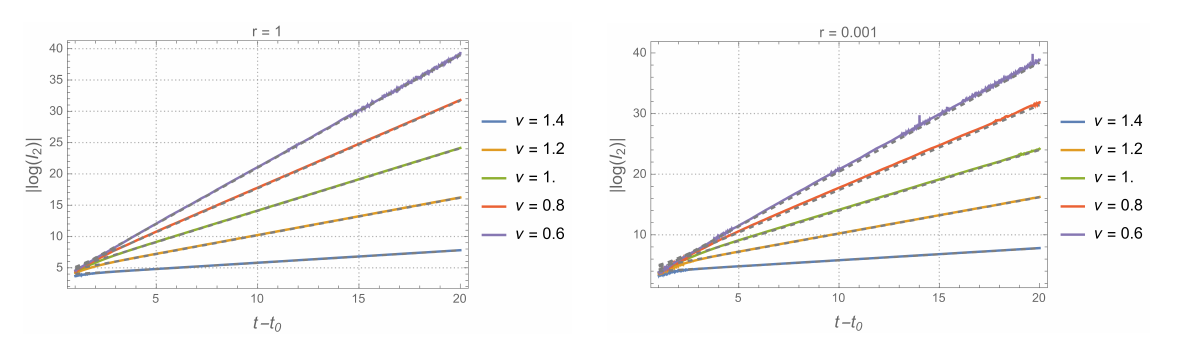}
    \caption{The numerical plots of $\left|\log(I_2)\right|$ as a function of time $t-t_0$ for masses $m<\sqrt{2}$, $r=1$ (left) and $r=0.001$ (right). The dashed lines show the analytical estimate. We chose $-t_0=\log(102)\simeq 4.6.$}
    \label{Fig:subcritical_masses}
    \hfill
\end{figure}

\begin{figure}[t]
    \centering
    \def\svgwidth{\textwidth}
    \includegraphics[width=\linewidth]{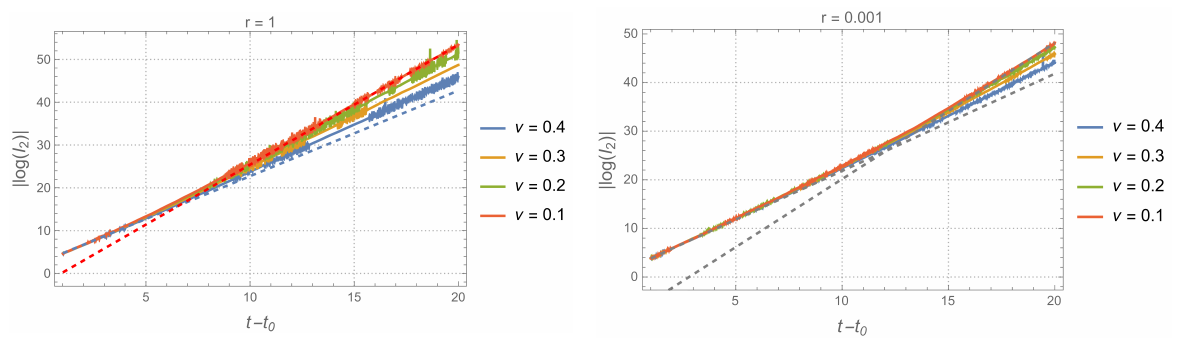}
    \caption{The numerical plot of $\left|\log(I_2)\right|$ as a function of time $t-t_0$ for masses $m>\sqrt{2}$. Here we see that up to $t\lesssim -\log(r)$ the slope equals $2<3-2\nu$ (blue dashed line), independently of the mass, and then for $t>-\log(r)$ it matches the slope $3-2\nu$ (red dashed line for $\nu=0.1$).}
    \label{Fig:supercritical_masses}
    \hfill
\end{figure}

An interesting feature of the modified (quenched) Keldysh function (\ref{eq:EPP_general_keldysh_Q}) is that for $m<m_{cr}=\sqrt{2}$ ($\nu>\nu_{cr}=\frac{1}{2}$) the decay proceeds with the power $3-2\nu$, whereas for $m>m_{cr}$ the leading decay power equals $2<3-2\nu$ for times $t\lesssim-\log(r)$ (or $\eta\gtrsim r$). This follows from (\ref{eq:EPP_I2_estimation}), since in that case one can approximately replace $\sin\Big(\ell\frac{r}{\eta_0}\Big)$ by its argument over the entire integration interval. For such $m>m_{cr}$ from the complementary series this may be important for the evolution of local operators in the limit $r\rightarrow 0$ after the subtraction of UV divergences, because the decay of the correction is slower than that of the Bunch–Davies correlator, $iG_{K} \sim\left(\frac{\eta}{r}\right)^{3-2\nu}$, which reaches this asymptotic behavior rather quickly even for $r<\eta$. This circumstance may spoil its “attractor” property in such a situation. Although fields with $m>\sqrt{2}$ are of limited practical relevance for inflationary physics, they remain interesting from a theoretical standpoint, and the full spectrum of primordial fields has yet to be determined. The above observation for the asymptotic behavior of the integral $I_2^{\text{EPP}}$ can be checked numerically using the exact formula (\ref{eq:I2_general_EPP}) for $m<m_{cr}$ and $m>m_{cr}$ (see fig.\ref{Fig:subcritical_masses} and fig.\ref{Fig:supercritical_masses}).

In the case $m=\sqrt{2}$ we have $2=3-2\nu$, and the integrals for the nonequilibrium Keldysh propagator can be evaluated exactly:
\begin{equation}\label{eq:confromal_4d_postQ_Keldysh}
    \begin{aligned}
        iG^Q_K(t,\mathbf{x}|t,\mathbf{y}) = \frac{1}{4\pi^2}\frac{\eta^2}{r^2} + \frac{1}{2\pi^2}\cdot\left(\frac{\eta}{\eta_0}\right)^2\cdot\Bigg\{-\alpha \frac{\eta_0}{2r}\log\left[\frac{2(\eta_0-\eta)+r}{2(\eta_0-\eta)-r}\right] \; + 
        \\
        +\;\alpha^2\log\left[\frac{\sqrt{4(\eta_0-\eta)^2-r^2}}{r}\right] + \alpha^2\frac{\eta_0-\eta}{r}\log\left[\frac{2(\eta_0-\eta)+r}{2(\eta_0-\eta)-r}\right]\Bigg\}.
    \end{aligned}
\end{equation}

\begin{figure}[t]
    \centering
    \def\svgwidth{\textwidth}
    \includegraphics[width=\linewidth]{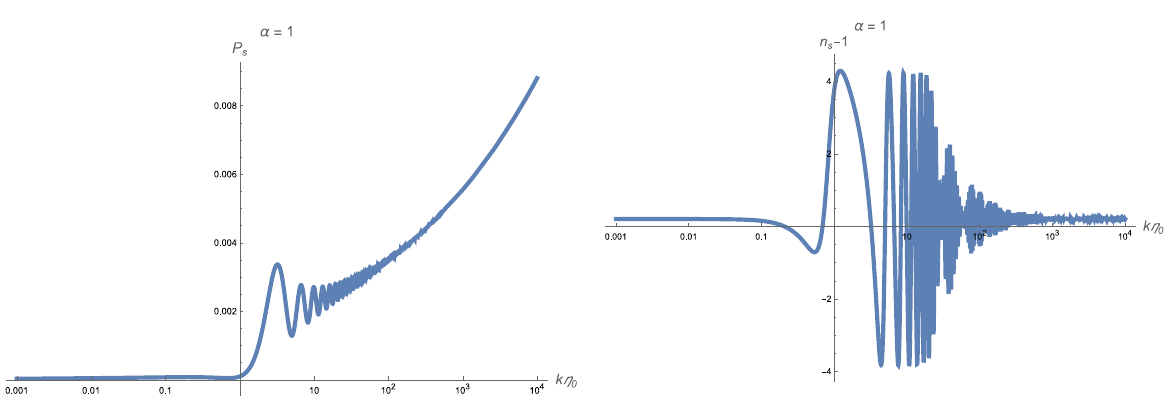}    
    \caption{Power spectrum $\mathcal{P}_s$ and spectral index $n_s-1$ as the functions of the scale $k\eta_0$ for $\alpha=1$. We chose $\eta_0=1$ and $\eta=10^{-6}$ so that all depicted scales are super-horizon.}
    \label{Fig:PsNsA1}
    \hfill
\end{figure}

Another application of nonequilibrium dynamics on the de Sitter background that is relevant for the early Universe is the study of the scalar power spectrum \cite{Joy:2007na, Banerjee:2021lqu, Carrilho:2016con, Hazra:2010ve}. If, for example, our scalar field describes fluctuations of the inflaton, then for the power spectrum $\mathcal{P}_s(k)$ and the spectral index $n_s(k)$ we have:
\beq\label{eq:PsNs_definition}
    \mathcal{P}_s(k) = \frac{k^3}{2\pi^2}iG_{K}^{Q}(\bm{k}|t,t),\quad n_s(k)-1=\frac{d\log\mathcal{P}_s}{d\log k},
\eeq
where $iG_{K}^{Q}(\bm{k}|t,t)$ is simply the Fourier transformation of the two-point correlator (\ref{eq:EPP_general_keldysh_Q}). In our Gaussian-quench setup, the plots for $\alpha=1$ and a small mass ($\nu=1.4$) look as in fig.\ref{Fig:PsNsA1}. 

As seen from this figure, the spectral index features a sharp and comparatively broad first peak, followed by more frequent oscillations with decreasing amplitude. A similar pattern appears for the $m_{\text{eff}}(t)$ protocol for an inflaton potential \cite{Joy:2007na, Hazra:2010ve} and for the $O(N)$ model in de Sitter \cite{Carrilho:2016con}. We will return to cosmological observables for non-Gaussian initial-state perturbations, as well as to systems with a large number $N$ of fields, in future work.

\section{Comment on the non-gaussian quench dynamics}\label{sec:comment}
One of the key distinctions of our approach to describe relaxation in a non-equilibrium initial state is that analytical calculations are not restricted only to Gaussian distributions in the initial density matrix, but non-gaussianities can also be incorporated. In this section, we demonstrate how to correctly define such an initial state and apply the method in the two-dimensional case $D=2$ for the two-point correlator in flat spacetime.

Let us consider, for example, the case with the insertion of a cubic operator $\widehat{V}$ (with coupling ``constant'' $\lambda$) and the corresponding functional operator $Q$, defined in (\ref{eq:quench_functional_operator}):
\beq\label{eq:qubic_quench}
\begin{aligned}
    \widehat{V}\left[\widehat{\phi}\right] = \int d^{D-1} \bm{x} \sqrt{|g(t_{0},\bm{x})|}\; :\widehat{\phi}^3:(\bm{x})\;, 
    \\
    Q\left[\Phi, - \frac{\delta}{\delta \Pi}\right] = \exp{\left\{\lambda \int d^{D-1}\bm{x} \sqrt{|g(t_{0},\bm{x})|} \left[3 :\Phi^2:(\bm{x})\frac{\delta}{\delta \Pi(\bm{x})} - \frac{1}{4} \frac{\delta^3}{\delta \Pi^{3}(\bm{x})}\right]\right\}}\;.
\end{aligned}
\eeq
Here, normal ordering\footnote{In general, all additional terms in the Hamiltonian should be normal ordered $\delta\widehat{H} \sim \lambda :\widehat{V}\left[\widehat{\phi}\right]:\;$. In the Gaussian quench case the result does not depend on this, which is why we did not impose it earlier.} refers to the subtraction of the counterterm proportional to $\Delta(x,x) = iG_K(x,x)$, the two-point correlator evaluated at coincident points in some ultraviolet regularization scheme:
\beq\label{eq:normal_ordering}
    \begin{aligned}
        :\!\phi^2\!:(x) = \phi(x)^2 - \Delta(x,x),
            \\
        :\!\phi^3\!:(x) = \phi(x)^3 - 3\,\Delta(x,x)\,\phi(x).      
    \end{aligned}
\eeq
Following the procedure outlined in Appendix \ref{app:keldysh_formalism}, we obtain the expression for the two-point correlator after such a non-Gaussian disturbance:
\beq\label{eq:keldysh_non_gaussian}
    \begin{aligned}
        iG_{K}^{Q}(t,\bm{x}|t,\bm{y}) 
    =\Big\langle \widehat{\phi}(\bm{x})\widehat{\phi}(\bm{y})\Big\rangle_{t}^{Q} = 
    i G_{K}(t, \bm{x} | t, \bm{y})\; -
    \\
    - \; 9 \lambda^2 \int d^{D-1}\bm{\xi}_{1} d^{D-1} \bm{\xi}_{2} \sqrt{|g(t_{0},\bm{\xi}_{1})|} \sqrt{|g(t_{0},\bm{\xi}_{2})|}
    \cdot G_{R}(t, \bm{x}| t_{0}, \bm{\xi}_{1}) G_{R}(t, \bm{y}| t_{0}, \bm{\xi}_{2})G^2_{K}(t_{0}, \bm{\xi}_{1}| t_{0}, \bm{\xi}_{2})\;,
    \end{aligned}
\eeq
or, in momentum space, for the correction $i\Delta G_{K}$ to the tree-level Keldysh propagator $iG_{K}$:
\beq\label{eq:keldysh_nongaussian_momentum_space}
\begin{aligned}
    i \Delta G_{K}(t, \bm{x}| t, \bm{y}) = - 18 \lambda^2 \int \frac{d^{D-1}\bm{k}}{(2\pi)^{D-1}} \int \frac{d^{D-1}\bm{q}}{(2\pi)^{D-1}} \sqrt{|g(t_{0})|}\sqrt{|g(t_{0})|} e^{i \bm{k}(\bm{x}-\bm{y})} \; \times \\ \times \; G_{R}\left(\bm{k}|t, t_{0}\right) G_{R}\left(-\bm{k}|t, t_{0}\right) G_{K}\left(\frac{\bm{k}+\bm{q}}{2}\bigg|t, t_{0}\right) G_{K}\left(\frac{\bm{k}-\bm{q}}{2}\bigg|t, t_{0}\right).
\end{aligned}
\eeq
It is easy to see that this expression is essentially a loop integral over the momentum $\bm{q}$ and, in particular, contains ultraviolet divergences associated with such loops. Therefore, in the general case, instantaneous interaction counterterms at the initial Cauchy surface, $t=t_0$, with renormalized couplings must be included. For now, we postpone this issue and instead examine the two-dimensional flat space, $D=2$, where the integral in (\ref{eq:keldysh_nongaussian_momentum_space}) is finite. Then:
\beq\label{eq:nongausisian_keldysh_2}
    i \Delta G_{K}(t, \bm{x}| t, \bm{y}) = \frac{18 \lambda^2}{m^2} \int \frac{d \xi}{2\pi} \int \frac{d \zeta}{2\pi} e^{i \xi m(x-y)}\frac{1 - \cos{\left(2\sqrt{\xi^2 + 1}\;mt\right)}}{\xi^2 + 1} \frac{1}{\sqrt{(\xi^2-\zeta^2)^2 + 8(\xi+\zeta)+16}}.
\eeq
As in the Gaussian quench case in flat space, this integral contains a time-independent contribution and a time-dependent ‘‘tail’’ represented by the cosine term in (\ref{eq:nongausisian_keldysh_2}). The $\zeta$-integral can be performed exactly, and the time dependence of the tail may be estimated via the stationary phase method for $mt\gg1$, yielding:
\beq\label{eq:nongausisian_keldysh_estimation}
\begin{aligned}
    i \Delta G_{K}(t, \bm{x}| t, \bm{y}) \approx I - \frac{9 \lambda^2 \sqrt{\pi}}{2 m^{5/2} \sqrt{t}} \cos{\left(2 m t + \frac{\pi}{4}\right)},
    \\
    I = \frac{36 \lambda^2}{m^2} \int\limits_{-\infty}^{+\infty} \frac{d \xi}{2 \pi} \frac{e^{i\xi m(x-y)}}{|\xi|+2} \frac{1}{\xi^2 + 1} K\left(\frac{4 \sqrt{|\xi|}}{|\xi|+2}\right), 
\end{aligned}
\eeq
where $K(z)$ is the complete elliptic integral of the first kind. As we see, the relaxation rate at late times for the massive theory is the same as in the Gaussian case (see also Appendix \ref{app:flat}), and appears to be insensitive to the particular quench type, at least for the two-point function in a massive theory.

Nevertheless, from (\ref{eq:qubic_quench}) it is clear that such an initial state modifies higher-point correlators in a nontrivial way. In particular, one may study the evolution and relaxation of systems with nontrivial initial correlations encoded in the density matrix. Of special interest is the question, still not completely settled, of how initial correlations can imprint themselves during the inflationary period of the Universe, which can be understood by studying multi-point correlators of primordial fluctuations. We will return to the behavior of QFT in $D>2$ under non-Gaussian quenches in future work.

\section{Conclusion}\label{sec:discussion}
In this work, we examined the dynamics of quantum fields in a specific class of nonequilibrium initial states on curved space-time backgrounds. Developing the approach of \cite{Radovskaya:2023ctz}, we expressed the post-quench two-point correlation function in terms of the Wigner functional of the initial density matrix, with the quench implemented as a unitary transformation of the initial state: $\widehat{\rho}_Q=\widehat{Q}\widehat{\rho}_0\widehat{Q}^{\dagger}$.
This formulation, based on the Keldysh–Schwinger functional integral, allows one to clearly separate the modification of the initial state from the subsequent unitary evolution and is applicable to an arbitrary curved space-time geometry. As a result, the method is particularly useful in situations where the standard description via sudden Hamiltonian changes or explicit mode evolution becomes technically cumbersome or inefficient.

As a first example, we considered in detail the Gaussian (quadratic) quench, corresponding to an initial-state perturbation of the form (\ref{eq:qudratic_quench}). Such a quench modifies the initial statistical distribution rather than the Hamiltonian itself, and the post-quench dynamics can be expressed entirely through equilibrium Green’s functions. In this case, the Keldysh function after the quench $G_{K}^{Q}$ takes the compact form (\ref{eq:Keldysh_evolution_general}) and (\ref{eq:keldysh_evolution_momentum_space}), so that the effect of a global quench reduces to a finite number of correction terms added to the standard correlators.

We then illustrated the method in flat space-time, see Appendix \ref{app:flat}. For massive fields, we obtained relaxation of the correlator to a modified post-quench value with an oscillatory power-law tail, decaying at large times as shown in (\ref{eq:post_quench_Keldysh_flat_spacetime}). This behaviour matches the well-known results for mass-dependent quench protocols and therefore provides a nontrivial consistency check of the method. However, for massless fields in $D=4$ we found qualitatively different dynamics: the correlator does not relax even at late times, see (\ref{eq:flat_massless_post_quench_evolution}). This indicates that the class of nonequilibrium states considered here cannot in general be reduced to the standard mass-quench scenario. A more detailed analysis of this ``critical'' post-quench dynamics will be pursued in future work.

Our primary interest concerns the dynamics in de Sitter space, see Section \ref{sec:EPP}. For the Gaussian quench, we analyzed the late-time asymptotics of the two-point function. We found that the decay of the quench-induced corrections depends sensitively on the mass of the field: for light fields ($m<\sqrt{2}$, or equivalently $\nu>1/2$) the corrections decay as $(\eta/\eta_0)^{3-2\nu}$, whereas heavier fields exhibit a slower decay regime in the limit of close spatial points. This suggests that post-quench dynamics of local observables such as the stress-energy tensor may depend nontrivially on the mass spectrum, which is relevant for discussions of the stability of the Bunch–Davies state and for understanding secular (time-growing) quantum corrections in de Sitter space. In addition, we analyzed the possible implications of this type of post-quench description for cosmological observables in the expanding Poincaré patch. In particular, we computed the power spectrum $\mathcal{P}_s(k)$ and the spectral index $n_s(k)$ according to (\ref{eq:PsNs_definition}). The resulting spectra exhibit a pronounced initial peak followed by decaying oscillations (see Fig.~\ref{Fig:PsNsA1}). Such features are qualitatively similar to those encountered in models with time-dependent effective mass $m_{\text{eff}}(t)$ \cite{Joy:2007na,Hazra:2010ve} and in $O(N)$ models in de Sitter space \cite{Carrilho:2016con}. This should not be regarded as an alternative to these approaches. Instead, our point is that nontrivial initial state distributions should also be incorporated in a full and consistent treatment of the problem.

Finally, in Section \ref{sec:comment} we showed that the method naturally generalizes to non-Gaussian initial states generated by higher-order operators, using the cubic perturbation $\widehat{V}\sim\int:\phi^3:$ as an example. In the simplest case of $D=2$ Minkowski spacetime, we found contributions that also decay at late times with a power-law behaviour. This demonstrates the applicability of the method to systematic studies of non-Gaussianity and higher-order correlation functions, including in cosmological settings where primordial non-Gaussianities may play a role.

Future directions include analyzing the impact of such nonequilibrium initial states on secular loop corrections in de Sitter space, as well as computing the behaviour of the energy–momentum tensor and other local observables after proper regularization in the coincident limit. The approach developed here provides a convenient and analytically controlled framework for these investigations.

\section*{Acknowledgments}
We would like to acknowledge fruitful discussions with Dmitry S. Ageev and A.G. Semenov. We are particularly grateful to E.T. Akhmedov for valuable discussions and careful reading of the paper. The work of Damir Sadekov was supported by the Foundation for the Advancement of Theoretical Physics and Mathematics ``BASIS''.

\appendix

\section{Quantum quench in curved space-time}\label{app:keldysh_formalism}
Consider a free massive minimally coupled scalar field $\phi$ in curved space-time:
\beq\label{eq:free_field_action_curved_spacetime}
    S[\phi] =  \frac{1}{2} \int_{t_{0}}^{\infty} d t\int d^{D-1} \bm{x} \sqrt{|g(t,\bm{x})|} \Big[\partial_{\mu} \phi \partial^{\mu} \phi - m^2 \phi^2 \Big]. 
\eeq
Then the time evolution of the expectation value of a composite operator $\widehat{O}\big[\widehat{\phi}(\bm{x})\big]$ is given by
\beq\label{eq:basic_operator_evolution}
    \Big\langle \widehat{O}\big[\widehat{\phi}(\bm{x})\big] \Big\rangle_{t} \equiv \text{tr}\Big\{\widehat{\rho}(t)\widehat{O}\Big\} = \text{tr}\Big\{\widehat{\rho}\widehat{U}^{\dagger}(t,t_0)\widehat{O}\widehat{U}(t,t_0) \Big\} \equiv \text{tr}\Big\{\widehat{\rho}\cdot\widehat{U}^{\dagger}(t,t_0)\widehat{O}\widehat{U}^{\dagger}(+\infty,t)\cdot\widehat{U}(+\infty,t_0)\Big\},
\eeq
where $\widehat{\rho}$ is the initial density matrix and $\widehat{U}(t,t_0)$ is the evolution operator. Expanding the trace as an integral over the basis of field eigenstates $\big|\xi_1\big\rangle$, and inserting the corresponding resolutions of the identity at the marked points in (\ref{eq:basic_operator_evolution}), as well as using the standard transition from the evolution operator to the path integral, we obtain the expression
\beq\label{eq:basic_operator_evolution_func_int}
\begin{aligned}
    \Big\langle \widehat{O}\big[\widehat{\phi}(\bm{x})\big] \Big\rangle_{t} = 
    \\
    =\int \mathscr{D}\chi(\bm{x}) \mathscr{D}\xi_1(\bm{x}) \mathscr{D}\xi_2(\bm{x})\Big\langle\xi_1\Big|\widehat{\rho}\Big|\xi_{2}\Big\rangle \int_{\phi_{+}(t_0)=\xi_1}^{\phi_{+}(+\infty)=\chi}\mathscr{D}\phi_{+}\int_{\phi_{-}(t_0)=\xi_2}^{\phi_{-}(+\infty)=\chi}\mathscr{D}\phi_{-}e^{iS[\phi_{+}]-iS[\phi_{-}]}\cdot O\Big[\phi_{-}(t,\bm{x})\Big].
\end{aligned}
\eeq
In the Schwinger–Keldysh functional integral formalism it is convenient to introduce the variables
\[
\phi_{cl} = \frac{1}{2}\left(\phi_{+} + \phi_{-}\right),\quad \phi_{q} = \phi_{+} - \phi_{-}.
\]
Then the action on the Keldysh time contour takes the form
\beq\label{eq:Keldysh_action}
\begin{aligned}
    S_K[\phi_{cl},\phi_{q}] = S[\phi_{+}]-S[\phi_{-}] =   \int_{t_{0}}^{\infty} dt \int d^{D-1} \bm{x} \sqrt{|g|} \Big[g^{\mu\nu}\partial_{\mu} \phi_{q} \partial_{\nu} \phi_{cl} + m^2 \phi_{q} \phi_{cl}\Big] = 
    \\
    =  -\int d^{D-1}\bm{x}\sqrt{|g|}\cdot\Big(\xi_1(\bm{x})-\xi_2(\bm{x})\Big)\cdot g^{\mu0}\pd_{\mu}\phi_{cl}(t_0,\bm{x}) - \int d^{D}x\sqrt{|g|}\;\phi_q \left[\frac{1}{\sqrt{|g|}}\pd_{\mu}\sqrt{|g|}g^{\mu\nu}\pd_{\nu} + m^2\right]\phi_{cl} = 
    \\
    =-\int d^{D-1}\bm{x}\cdot\Big(\xi_1(\bm{x})-\xi_2(\bm{x})\Big)\cdot\pi(t_0,\bm{x}) - \int d^{D}x \sqrt{|g|}\;\phi_{q}\cdot\text{EOM}\Big[\phi_{cl}\Big],
\end{aligned}   
\eeq
where we have introduced the canonical momentum $\pi(t,\bm{x}) = \sqrt{|g|}g^{00}\pd_{0}\phi(t,\bm{x})$, and denoted the operator in the square brackets in the second line acting on $\phi_{cl}$ as $\text{EOM}\big[\phi_{cl}\big]$. We also assumed that the mixed components $g^{i0}$ vanish for spatial index $i$. Using the further convenient notations $\beta(\bm{x})=\xi_1(\bm{x})-\xi_2(\bm{x})$ and $\Phi(\bm{x})=\frac{1}{2}\big(\xi_1(\bm{x})+\xi_{2}(\bm{x})\big)$, we can rewrite (\ref{eq:basic_operator_evolution_func_int}) in the form
\beq\label{eq:operator_evolution_wigner_function}
\begin{aligned}
     \Big\langle \widehat{O}\big[\widehat{\phi}(\bm{x})\big] \Big\rangle_{t} = \int \mathscr{D}\chi(\bm{x}) \mathscr{D}\Phi(\bm{x})\mathscr{D}\beta(\bm{x})\bigg\langle\Phi+\frac{\beta}{2}\bigg|\widehat{\rho}\bigg|\Phi-\frac{\beta}{2}\bigg\rangle\times
     \\
     \times \int_{\phi_{cl}(t_0)=\Phi}^{\phi_{cl}=\chi}\mathscr{D}\phi_{cl}\int_{\phi_q(t_0)=\beta}^{\phi_q=0}\mathscr{D}\phi_{q}\;\exp\left\{-i\int d^{D-1}\bm{x}\beta(\bm{x})\pi(t_0,\bm{x})\right\}\times
     \\
     \times\exp\left\{i\int d^Dx\sqrt{|g|}\phi_{q}\cdot\text{EOM}[\phi_{cl}]\right\}\;O\bigg[\phi_{cl}-\frac{\phi_q}{2}\bigg]=
     \\
     = \int \mathscr{D}\Phi(\bm{x})\mathscr{D}\Pi(\bm{x})\mathscr{D}\beta(\bm{x})\int_{\phi_{cl}=\Phi}\mathscr{D}\phi_{cl}\exp\left\{-i\int d^{D-1}\bm{x}\beta(\bm{x})\Pi(\bm{x})\right\}\times
     \\
     \times\delta\Big(\Pi(\bm{x})-\pi(t_0,\bm{x})\Big)\cdot\delta\Big(\text{EOM}[\phi_{cl}]\Big)\cdot O[\phi_{cl}]
     =\int \mathscr{D}\Phi(\bm{x})\mathscr{D}\Pi(\bm{x})W\left[\Phi,\Pi\right]\cdot O\Big[\phi\Big|_{\text{on-shell}}\Big],
\end{aligned}
\eeq
where in the fourth line we have omitted $\phi_q$ inside the composite operator $O$, since the contraction $\wick{\c\phi_q\c \phi_q}=0$ in the Schwinger–Keldysh technique and the mixed contractions $\wick{\c\phi_q\c\phi_{cl}}$ vanish at coincident times due to causality (see also \cite{Leonidov:2014dfa}). Also, in the fourth line, we integrated over $\phi_q$, producing the delta-function, which enforces $\text{EOM}[\phi_{cl}]=0$, and introduced the field $\Pi(\bm{x})$ through the additional delta-function. In the last line, we introduced the Wigner functional and carried out the integration over $\phi_{cl}$ using the delta-constraints, which project $\phi_{cl}$ onto the solutions of the equation of motion with the specified initial conditions:
\beq\label{eq:wigner_func_def}
    W\left[\Phi,\Pi\right] = \int \mathscr{D}\beta(\bm{x})\exp\left\{-i\int d^{D-1}\bm{x}\;\beta(\bm{x})\pi(\bm{x})\right\}\Big\langle\Phi+\frac{\beta}{2}\Big|\widehat{\rho}\Big|\Phi-\frac{\beta}{2}\Big\rangle;
\eeq
\beq\label{eq:on_shell_conditions}
    \text{EOM}\left[\phi\Big|_{\text{on-shell}}\right] = 0,\;\;\phi(t_0,\bm{x})\Big|_{\text{on-shell}} = \Phi(\bm{x}),\;\;\pd_{t}\phi(t_0,\bm{x})\Big|_{\text{on-shell}} = \frac{1}{\sqrt{|g|}g^{00}}\Pi(\bm{x})\;.
\eeq
If one can find the retarded Green function for the given gravitational background
\beq\label{eq:retarded_Green_function_def}
    \left[\frac{1}{\sqrt{|g|}}\pd_{\mu}\sqrt{|g|}g^{\mu\nu}\pd_{\nu} + m^2\right]G_{R}\left(t, \bm{x}| t', \bm{y}\right) = - \frac{\delta(t-t')\delta^{(3)}(\bm{x-\bm{y}})}{\sqrt{|g|}},
\eeq
then the on-shell field is expressed explicitly as
\beq\label{eq:solution_on_shell_general}
    \phi(t,\bm{x})\Big|_{\text{on-shell}} = -\int d^{D-1}\bm{y}\bigg[\sqrt{|g(t_0,\bm{y})|}\;g^{00}(t_0,\bm{y})\;\pd_{t}G_{R}\left(t,\bm{x}|t_0,\bm{y}\right)\cdot\Phi(\bm{y}) + G_{R}\left(t,\bm{x}|t_0,\bm{y}\right)\cdot\Pi(\bm{y}) \bigg].
\eeq

Now we want to study the dynamics of correlation functions of local operators after the system is driven out of an equilibrium state described by some initial density matrix $\widehat{\rho}_0$. If, at time $t_0$, the Hamiltonian $\widehat{H}$ of the system undergoes a perturbation $\delta\widehat{H} = \alpha\,\delta(t-t_0)\widehat{V}[\widehat{\phi}(\bm{x})]$, then the density matrix describing the subsequent evolution becomes
\beq\label{eq:density_matrix_change_after_quench}
    \widehat{\rho}_{Q}(t_0) = \widehat{Q}\widehat{\rho}_{0}\widehat{Q}^{\dagger}, \;\;\widehat{Q} = e^{-i\alpha \widehat{V}[\widehat{\phi}(\bm{x})]}. 
\eeq
More generally, we can simply consider perturbed initial states of the form (\ref{eq:density_matrix_change_after_quench}) without referring to the Hamiltonian quench interpretation. Using the identity
\beq\label{eq:pi_derivative_exponent}
    -i\frac{\delta}{\delta \Pi(\bm{y})}\exp\left\{i\int d^{D-1}\bm{x}\beta(\bm{x})\Pi(\bm{x})\right\} = \beta(\bm{y})\exp\left\{i\int d^{D-1}\bm{x}\beta(\bm{x})\Pi(\bm{x})\right\},
\eeq
the modified Wigner functional (\ref{eq:wigner_func_def}) after the quench takes the form
\beq\label{eq:modified_wigner_functional}
\begin{aligned}
    W_{Q}\left[\Phi,\Pi\right] =  \int \mathscr{D}\beta(\bm{x})\exp\left\{-i\int d^{D-1}\bm{x}\;\beta(\bm{x})\pi(\bm{x})\right\}\Big\langle\Phi+\frac{\beta}{2}\Big| \widehat{Q}\,\widehat{\rho}_{0}\,\widehat{Q}^{\dagger}\Big|\Phi-\frac{\beta}{2}\Big\rangle = 
    \\
    = Q\left[\Phi, \frac{\delta}{\delta\Pi}\right]W_{0}\left[\Phi,\Pi\right],
\end{aligned}
\eeq
where $W_0[\Phi,\Pi]$ corresponds to the unperturbed density matrix $\widehat{\rho}_0$, and we introduced the functional operator
\beq\label{eq:quench_functional_operator}
    Q\left[\Phi, \frac{\delta}{\delta\Pi}\right] = \exp\left\{-i\alpha\left(V\left[\Phi-\frac{i}{2}\frac{\delta}{\delta \Pi}\right] - V\left[\Phi+\frac{i}{2}\frac{\delta}{\delta \Pi}\right]\right)\right\}.
\eeq
Substituting (\ref{eq:modified_wigner_functional}) into (\ref{eq:operator_evolution_wigner_function}) and integrating by parts over $\Pi(\bm{x})$, we obtain the general post-quench evolution:
\beq\label{eq:operator_evolution_after_quench}
    \Big\langle \widehat{O}\big[\widehat{\phi}(\bm{x})\big] \Big\rangle_{t}^{Q} = \int \mathscr{D}\Phi(\bm{x})\mathscr{D}\Pi(\bm{x})\,W_{0}\left[\Phi,\Pi\right]\;Q\left[\Phi, -\frac{\delta}{\delta\Pi}\right]O\Big[\phi\Big|_{\text{on-shell}}\Big].
\eeq
Thus, the problem of post-quench evolution reduces to computing expectation values with the equilibrium density matrix, but with the observable modified by the functional operator $Q$.

As a simple example of a global quench, consider a Gaussian initial state, which can be interpreted as an instantaneous perturbation of the mass term:
\beq\label{eq:qudratic_quench}
\begin{aligned}
    \widehat{V}\left[\widehat{\phi}\right] = \int d^{D-1}\bm{x}\sqrt{|g(t_0,\bm{x})|}\;\widehat{\phi}^2(\bm{x})\;,
    \\
    Q\left[\Phi,-\frac{\delta}{\delta\Pi}\right] = \exp\left\{2\alpha\int d^{D-1}\bm{x}\;\Phi(\bm{x})\frac{\delta}{\delta\Pi(\bm{x})}\right\}.
\end{aligned}
\eeq
Finally, using (\ref{eq:solution_on_shell_general}) and the representation of the Keldysh propagator in terms of the Wigner functional,
\beq\label{eq:Keldysh_through_func_integral}
    iG_K(t,\bm{x}|t',\bm{y}) = \frac{1}{2}\text{tr}\Big(\widehat{\rho}\cdot\left\{\widehat{\phi}(t,\bm{x}),\widehat{\phi}(t',\bm{y})\right\}\Big) = \int \mathscr{D}\Phi(\bm{x})\mathscr{D}\Pi(\bm{x})\;W\left[\Phi,\Pi\right]\phi(t,\bm{x})\Big|_{\text{on-shell}}\phi(t',\bm{y})\Big|_{\text{on-shell}}, 
\eeq
we obtain the evolution of the equal-time two-point correlator:
\beq\label{eq:two_point_evolution_general}
\begin{aligned}
    iG_{K}^{Q}(t,\bm{x}|t,\bm{y}) 
    =\Big\langle \widehat{\phi}(\bm{x})\widehat{\phi}(\bm{y})\Big\rangle_{t}^{Q} = 
    \\    =\int\mathscr{D}\Phi(\bm{x})\mathscr{D}\Pi(\bm{x})W_0\left[\Phi,\Pi\right]\exp\left\{2\alpha\int d^{D-1}\bm{\xi}\sqrt{|g(t_0,\bm{\xi})|} \Phi(\bm{\xi})\frac{\delta}{\delta\Pi(\bm{\xi})}\right\}\phi(t,\bm{x})\Big|_{\text{on-shell}}\phi(t,\bm{y})\Big|_{\text{on-shell}} = 
    \\
    = iG_K(t,\bm{x}|t,\bm{y}) - 2\alpha\int d^{D-1}\bm{\xi}\sqrt{|g(t_0,\bm{\xi})|}\bigg[iG_K(t,\bm{x}|t_0,\bm{\xi})G_R(t,\bm{y}|t_0,\bm{\xi}) + iG_{K}(t,\bm{y}|t_0,\bm{\xi})G_R(t,\bm{x}|t_0,\bm{\xi}) \bigg] 
    \\
    +\; 4\alpha^2 \int d^{D-1}\bm{\xi}_1 d^{D-1}\bm{\xi}_2 \sqrt{|g(t_0,\bm{\xi}_1)|}\sqrt{|g(t_0,\bm{\xi}_2)|}\;iG_K(t_0,\bm{\xi}_1|t_0,\bm{\xi}_2)G_R(t,\bm{x}|t_0,\bm{\xi}_1)G_R(t,\bm{x}|t_0,\bm{\xi}_2).
\end{aligned}
\eeq
It is often convenient to evaluate these integrals in momentum space. If the background metric is homogeneous and isotropic, depending only on time, and the equilibrium state $\widehat{\rho}_0$ shares these symmetries, then the result takes the form
\beq\label{eq:two_point_evolution_momentum_space}
\begin{aligned}
    iG_{K}^{Q}(t,\bm{x}|t,\bm{y}) = \int\frac{d^{D-1}\bm{k}}{(2\pi)^{D-1}}\;e^{i\bm{k}(\bm{x-y})}\;\bigg\{iG_K(\bm{k}|t,t) \;-\;4\alpha\sqrt{|g(t_0)|}\;iG_{K}(\bm{k}|t,t_0)G_{R}(\bm{k}|t,t_0)\;+
    \\
    +\;4\alpha^2\sqrt{|g(t_0)|}\sqrt{|g(t_0)|}G_{K}(\bm{k}|t_0,t_0)G_{R}(\bm{k}|t,t_0)G_{R}(\bm{k}|t,t_0)\bigg\}.
\end{aligned}
\eeq
Note that for expectation values of polynomial operators we always obtain only a finite number of terms modifying the result relative to the unperturbed initial state. Moreover, expressions of the type (\ref{eq:two_point_evolution_general}) can be derived for higher-order correlation functions, as well as for non-Gaussian (non-quadratic) initial perturbations or spatially inhomogeneous quenches. Indeed, all arguments above remain valid if one replaces the constant $\alpha$ by a spatial function $\alpha(\bm{x})$, which simply appears inside the corresponding integrals.

\section{Quench dynamics in flat space-time}\label{app:flat}
Field quantization:
\beq\label{eq:field_quantization_flat_space}
    \widehat{\phi}(t,\bm{x}) = \int\frac{d^{D-1}\bm{k}}{(2\pi)^{D-1}}\Big[\widehat{a}_{\bm{k}}f_{k}(t) + \widehat{a}^{\dagger}_{\bm{k}}f_{k}^{*}(t)  \Big],
\eeq
where
\beq\label{eq:modes_flat_space}
\begin{aligned}
    \left[\widehat{a}_{\bm{p}},\widehat{a}^{\dagger}_{\bm{k}}\right] = (2\pi)^{D-1}\delta(\bm{k}-\bm{p}),
    \\
    f_{k}(t) = \frac{1}{\sqrt{2\sqrt{k^2+m^2}}}e^{-i\sqrt{k^2+m^2}t}.
\end{aligned}
\eeq
The corresponding propagators for the vacuum initial state are
\beq\label{eq:flat_space_propagators}
\begin{aligned}
    iG_{K}(\bm{k}|t,t_0) = \text{Re}\Big\{f_{k}(t)f_{k}^{*}(t_0) \Big\} = \frac{1}{2\sqrt{k^2+m^2}}\cos\Big(\sqrt{k^2+m^2}(t-t_0)\Big),
    \\
    G_{R}(\bm{k}|t,t_0) = -2\,\theta(t-t_0)\,\text{Im}\Big\{f_{k}(t)f_{k}^{*}(t_0) \Big\} = -\frac{1}{\sqrt{k^2+m^2}}\sin\Big(\sqrt{k^2+m^2}(t-t_0)\Big)\,\theta(t-t_0).
\end{aligned}
\eeq
In what follows we denote $r = |\bm{x}-\bm{y}|$ and assume $t>t_0$. Then, in flat spacetime, we can rewrite (\ref{eq:two_point_evolution_momentum_space}) as
\beq\label{eq:flat_quench_correction_general}
\begin{aligned}
    iG_{K}^{Q}(t,\bm{x}|t,\bm{y}) = iG_{K}(t,\bm{x}|t,\bm{y}) - 4\alpha\, I_{1}^{\text{flat}}(t,t_0,r) + 4\alpha^2\, I_2^{ \text{flat}}(t,t_0,r),
    \\
    I_{1}^{\text{flat}}(t,t_0,r) = -\frac{1}{4}\int\frac{d^{D-1}\bm{k}}{(2\pi)^{D-1}}\,e^{i\bm{k}(\bm{x}-\bm{y})}\frac{1}{k^2+m^2}\sin\Big(2\sqrt{k^2+m^2}(t-t_0)\Big),
    \\
    I_{2}^{\text{flat}}(t,t_0,r) = \frac{1}{2}\int\frac{d^{D-1}\bm{k}}{(2\pi)^{D-1}}\,e^{i\bm{k}(\bm{x}-\bm{y})} \frac{1}{\left(k^2+m^2\right)^{\frac{3}{2}}}\sin^{2}\Big(\sqrt{k^2+m^2}(t-t_0)\Big). 
\end{aligned}
\eeq
Using the Fourier-Bessel representation
\beq\label{eq:fourier_through_bessel}
    \int\frac{d^{D-1}\bm{p}}{(2\pi)^{D-1}}e^{i\bm{p}\bm{r}}f(p) = 
    \frac{1}{(2\pi)^{\frac{D-1}{2}}r^{\frac{D-3}{2}}}\int_{0}^{+\infty}dp\,p^{\frac{D-1}{2}}J_{\frac{D-3}{2}}(pr)f(p),
\eeq
for $D=4$, and setting $t_0=0$ due to time-translation symmetry, the integrals in (\ref{eq:flat_quench_correction_general}) become
\beq\label{eq:flat_integrals}
\begin{aligned}
    I_{1,4D}^{\text{flat}}(t,t_0,r) = -\frac{1}{8\pi^2 r}\int_{0}^{\infty}d\xi\frac{\xi}{\xi^2+1}\sin\Big(mr\cdot\xi\Big)\sin\Big(2mt\cdot\sqrt{\xi^2+1}\Big)
    \\
    I_{2,4D}^{\text{flat}} = \frac{1}{8\pi^2}\frac{1}{mr}\int_{0}^{\infty}d\xi\frac{\xi}{\left(\xi^2+1\right)^{\frac{3}{2}}}\sin\Big(mr\cdot\xi\Big)\bigg[1 - \cos\Big(2mt\cdot\sqrt{\xi^2+1}\Big)\bigg].
\end{aligned}
\eeq
We are interested in the late-time regime $t-t_0\gg m^{-1}$ and distances $r\ll t-t_0$. In this limit, the contributions from $I_1$ and the oscillatory part of $I_2$ are suppressed and can be estimated via the method of steepest descent. The non-oscillatory part of $I_2$ survives and yields the post-quench correction to the Keldysh function:
\beq\label{eq:post_quench_Keldysh_flat_spacetime}
\begin{aligned}
    iG_{K}^{Q}(t,\bm{x}|t,\bm{y}) = iG_{K}^{\text{post-quench}}(t,\bm{x}|t,\bm{y}) + \text{tail}(t,r),
    \\
    iG_{K}^{\text{post-quench}}(t,\bm{x}|t,\bm{y}) = \frac{1}{4\pi^2}\frac{m}{r}K_1(mr) + \frac{\alpha^2}{2\pi^2}K_0(mr),
    \\
    \text{tail}(t,r) \simeq \frac{\alpha m}{8\pi^{3/2}}\frac{1}{(mt)^{3/2}}\cos\Big(2mt+\frac{\pi}{4}\Big) 
    + \frac{\alpha^2}{8\pi^{3/2}}\frac{1}{(mt)^{3/2}}\sin\Big(2mt+\frac{\pi}{4}\Big)
    \sim \frac{1}{(mt)^{3/2}}\cos\big(2mt+\delta\big),
\end{aligned}
\eeq
with some phase $\delta$, defined in (\ref{eq:post_quench_Keldysh_flat_spacetime_Intro}). The relaxation rate coincides with the well-studied mass-quench dynamics in flat space \cite{Sotiriadis:2010si, Banerjee:2019ilw}.

\begin{figure}[t]
    \centering
    \includegraphics[width=\linewidth]{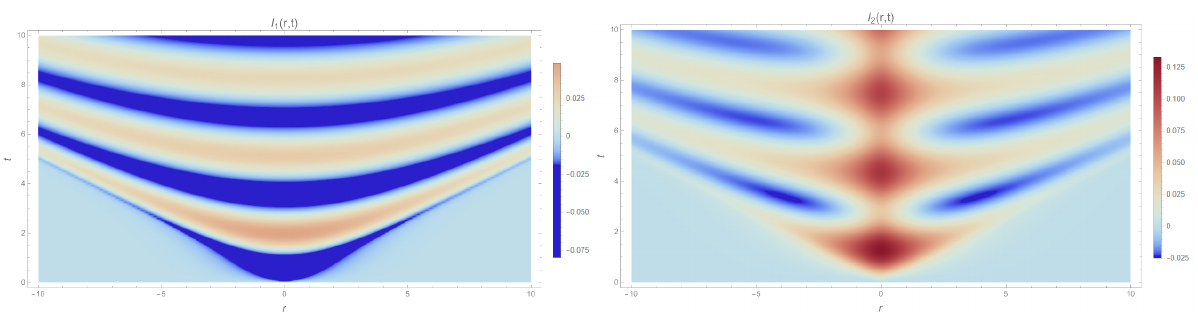}
    \caption{Numerical values of the integrals $I_1(t,r)$ and $I_2(t,r)$ in the case $D=2$. The ``horizon effect'' at $|r|<2t$ is clearly visible.}
    \label{Fig:Flat2D}
\end{figure}

The same can be done in two dimensions $(D=2)$:
\beq\label{eq:flat_integrals_2D}
\begin{aligned}
    I_{1,2D}^{\text{flat}}(t,t_0,r) = -\frac{1}{4\pi m}\int_{0}^{\infty}d\xi\frac{1}{\xi^2+1}\cos\Big(mr\cdot\xi\Big)\sin\Big(2mt\cdot\sqrt{\xi^2+1}\Big),
    \\
    I_{2,2D}^{\text{flat}} = \frac{1}{4\pi m^2}\int_{0}^{\infty}d\xi\frac{1}{\left(\xi^2+1\right)^{\frac{3}{2}}}\cos\Big(mr\cdot\xi\Big)\bigg[1 - \cos\Big(2mt\cdot\sqrt{\xi^2+1}\Big)\bigg],\quad r=x-y.
\end{aligned}
\eeq
In this case, the relaxation tail behaves as $\text{tail}(t)\sim\frac{1}{\sqrt{mt}}\cos\big(2mt+\delta_{2d}\big)$, consistent with the general expectations in even spatial dimensions. In addition, no subleading UV divergences appear, and although the integrals cannot be obtained in closed form, they can be computed numerically (see Fig.\ref{Fig:Flat2D}). As expected, the result for massive fields is quantitatively identical to the standard $m(t)$-quench protocol, including the ``horizon effect'' at $|r|>2t$ \cite{Sotiriadis:2010si}.

The situation changes drastically in the massless case $m=0$, where late-time suppression does not occur. However, in this case the post-quench dynamics (\ref{eq:flat_quench_correction_general}) can be computed exactly for $D=4$:
\beq\label{eq:flat_massless_post_quench_evolution}
    iG^{Q}_{K}(t,\bm{x}|t,\bm{y}) = \frac{1}{4\pi^2 r^2} + \frac{\alpha}{4\pi^2 r}\log\left[\frac{2t+r}{2t-r}\right]
    +  \frac{\alpha^2}{2\pi^2}\left\{\log\left[\frac{\sqrt{4t^2-r^2}}{r}\right] + \frac{t}{r}\log\left[\frac{2t+r}{2t-r}\right]\right\},
\eeq
which does not relax at $t\rightarrow\infty$ due to the first term in the braces. This type of ``non-relaxation'' behavior is also known in $D=2$ and $D=3$ \cite{Sotiriadis:2010si, Banerjee:2019ilw}, but does not occur in higher dimensions for a mass-quench. The origin and implications of this feature for our initial state will be discussed elsewhere. 

\bibliography{literature}
\bibliographystyle{unsrt}
\end{document}